\documentclass[traditabstract]{aa}

\pdfoutput=1
\usepackage{graphicx}
\usepackage[font=small]{caption}
\usepackage{indentfirst}
\usepackage{tabu}

\usepackage{natbib}
\usepackage{booktabs}
\usepackage{fixltx2e}

\usepackage{amsmath}
\usepackage{txfonts}
\usepackage{amssymb}
\usepackage[a]{esvect}
\usepackage{tikz}

\newcommand{\degree}[0]{$^{\circ}$}

\newcommand{\cbra}[1]{\left( #1 \right)}      

\let\baraccent=\=
\renewcommand{\=}[1]{\stackrel{#1}{=}} 
\let\arrowaccent=\>
\renewcommand{\>}[1]{\stackrel{#1}{\Rightarrow}} 

\makeatletter
\newcommand{\rmnum}[1]{{\footnotesize{\expandafter\@slowromancap\romannumeral #1@}}}
\newcommand{\Rmnum}[1]{{\expandafter\@slowromancap\romannumeral #1@}}
\makeatother

\everymath={\rm}
\newcommand{\tfm}[1]{\tablefootmark{#1}}
\begin{document}

\title{Jet-induced star formation in 3C 285 and Minkowski Object \thanks{Based on observations carried out with the IRAM 30m Telescope. IRAM is supported by INSU/CNRS (France), MPG (Germany) and IGN (Spain).}}

\author{
   Q. Salom\'e\inst{1} \and
   P. Salom\'e\inst{1} \and
   F. Combes\inst{1}
}

\institute{
   LERMA, Observatoire de Paris, CNRS UMR 8112, 61 avenue de l'Observatoire, 75014 Paris, France\\ email: quentin.salome@obspm.fr
}

\date{Received ??? / Accepted ??}

\titlerunning{Jet-induced star formation}
\authorrunning{Salom\'e Q. et al.}

\abstract{
   How efficiently star formation proceeds in galaxies is still an open question. Recent studies suggest that AGN can regulate the gas accretion and thus slow down star formation (negative feedback). However, evidence of AGN positive feedback has also been observed in a few radio galaxies (eg. Centaurus A, Minkowski Object, 3C 285, and the higher redshift 4C 41.17). \\
   Here we present CO observations of 3C 285 and Minkowski Object (MO), which are examples of jet-induced star formation. A spot (named 09.6) aligned with the 3C 285 radio jet, at a projected distance of $\sim 70 kpc$ from the galaxy centre, shows star formation, detected in optical emission. MO is located along the jet of NGC 541 and also shows star formation. To know the distribution of molecular gas along the jets is a way to study the physical processes at play in the AGN interaction with the intergalactic medium. \\
   We observed CO lines in 3C 285, NGC 541, 09.6 and MO with the IRAM-30m telescope. In the central galaxies, the spectra present a double-horn profile, typical of a rotation pattern, from which we are able to estimate the molecular gas density profile of the galaxy. The molecular gas appears to be in a compact reservoir, which could be an evidence of an early phase of the gas accretion after a recent merger event in 3C 285. In addition, no kinematic signature of a molecular outflow is detected by the 30m-telescope. \\
   Interestingly, the spot 09.6 and MO are not detected in CO. The cold gas mass upper limits are consistent with a star formation induced by the compression of dense ambient material by the jet. The depletion time scale in 09.6 and MO are of the order of and even smaller than what is found in 3C 285, NGC 541 and local spiral galaxies ($10^9\: yr$). The molecular gas surface density in the 09.6 spot follows a Schmidt-Kennicutt law if the emitting region is very compact as suggested by the $H\alpha$ emission, while MO is found to have a much higher SFE (very short depletion time). Higher sensitivity is necessary to detect CO in the spots of star formation, and higher spatial resolution is required to map the emission in these jet-induced star forming regions.}

\keywords{Methods:data analysis - Galaxies:evolution - interactions - star formation - Radio lines:galaxies}

\maketitle


\section{Introduction}

   The role played by AGN in galaxy evolution (and formation) has become a key question in the field of extragalactic astronomy in the past decade. The most studied phenomenon is the so-called AGN feedback that refers to a \textit{negative} feedback. In other words, the energy released by the AGN (mechanical or radiative) is supposed to be transferred to the surrounding medium and as a consequence to prevent or regulate star formation, either by heating or by expelling the gas reservoir available to fuel star formation (eg. \citealt{Fabian_2012,Heckman_2014} and references therein). However the details of how the AGN interacts with the gas of the host galaxy is still unclear. Another mechanism often referred to as AGN \textit{positive} feedback is also possible. The statistical study over hundreds of AGN by \cite{Zinn_2013} showed that AGN with pronounced radio jets have much higher star formation rate than the purely X-ray-selected ones. Supported by morphological association of AGN-jets and star forming regions \citep{Best_2012,Ivison_2012}, it is expected that propagation of shocks generated by the jets can accelerate gas cooling and thus trigger star formation. Such radio-jet/star formation associations were observed along radio-jets of local brightest cluster galaxies \citep{McNamara_1993} and molecular gas was mapped along the radio jet of Abell 1795 central galaxy \citep{SalomeP_2004}. In the early Universe, \cite{Emonts_2014} searched for CO in 13 high-z radio galaxies with redshifts between 1.4 and 2.8. The authors found CO-jet alignment in several of their sources and discuss possible explanations among which jet-induced star formation/gas cooling. \cite{Klamer_2004} also discussed this interpretation for detections of molecular gas spatially/kinematically offset from the central galaxy (and preferentially aligned along the radio axis) in $z>3$ sources. Jet triggered star formation processes (AGN positive feedback) were also modelled with numerical simulations. \cite{Fragile_2004} used hydrodynamic simulations of radiative shock-cloud interactions to show that it is possible to cool very efficiently large fraction of gas masses along the shock propagation path. This was also discussed in details by \cite{Gaibler_2012}.

   However, as mentioned above, evidences of radio-jet/molecular gas interaction has been found in very few objects: (1) Centaurus A, where the jet is encountering gas in the shells along its way \citep{Schiminovich_1994,Charmandaris_2000}; (2) Minkowski Object \citep{vanBreugel_1985}; (3) 3C 285 \citep{vanBreugel_1993} and, (4) at z = 3.8, the radio source 4C 41.17 \citep{Bicknell_2000,deBreuck_2005,Papadopoulos_2005}. Some other systems are suspected, in cooling flow clusters, such as Abell 1795 \citep{McNamara_2002,SalomeP_2004}, or Perseus A \citep{SalomeP_2008,SalomeP_2011} for instance. To better understand the physical processes at play in the AGN interaction with the intergalactic medium and its impact on star formation, it is important to know the molecular gas distribution in these objects.

   3C 285 is a double-lobed powerful FR-\Rmnum{2} radio galaxy where both lobes have a complex filamentary structure. In the eastern radio lobe, there is a radio jet with unresolved radio knots \citep{vanBreugel_1993}. A slightly resolved object is located near the eastern radio jet (3C 285/09.6; \citealt{vanBreugel_1993}). 3C 285/09.6 is a small, kiloparsec-sized object where star formation seems to be triggered by the jet from the radio source 3C 285 \citep{vanBreugel_1993}. Table \ref{table:overview_3C285} summarises general properties of the radio galaxy 3C 285 and the 09.6 spot.

\begin{table}[h]
  \centering
  \footnotesize
  \caption{\label{table:overview_3C285} General properties of 3C 285 and the 09.6 spot}
  \begin{tabu}{X[l]X[c]X[c]}
    \hline \hline
    Source                                 &           3C 285            &         3C 285/09.6         \\ \hline
    z                                      &            \multicolumn{2}{c}{0.0794 \tfm{a}}             \\
    $D_A$ (Mpc)                            &                  \multicolumn{2}{c}{309}                  \\
    $D_L$ (Mpc)                            &                  \multicolumn{2}{c}{360}                  \\
    Scale (kpc/$''$)                       &                  \multicolumn{2}{c}{1.5}                  \\ \hline
    RA (J2000)                             &    $13^h 21^m 17^s.813$     &    $13^h 21^m 22^s.134$     \\
    Dec (J2000)                            &       $+42:35:15.38$        &       $+42:35:20.14$        \\
    $v_{Hel}\: (km.s^{-1})$                &       0 ($z=0.0794$)        &    $-132\pm 25$ \tfm{a}     \\
    $m_B$ (mag)                            &        16.86 \tfm{b}        &                             \\
    $L_{H\alpha}\: (erg.s^{-1})$           & $2.5\times 10^{41}$ \tfm{c} & $2.8\times 10^{40}$ \tfm{a} \\
    $SFR_{H\alpha}\: (M_\odot.yr^{-1})$    &         1.34 \tfm{d}        &        0.15 \tfm{d}         \\
    $L_{24\: \mu m}\: (erg.s^{-1})$        &     $1.1\times 10^{44}$     &     $3.8\times 10^{42}$     \\
    $SFR_{24\: \mu m}\: (M_\odot.yr^{-1})$ &            12.0 \tfm{e}     &        0.61 \tfm{e}         \\
    $L_{TIR}\: (L_\odot)$                  &    $1.32\times 10^{11}$     &      $<5.3\times 10^9$      \\
    $SFR_{TIR}\: (M_\odot.yr^{-1})$        &         19.7 \tfm{d}        &       $<0.79$ \tfm{d}       \\ \hline
  \end{tabu}
  \tablefoot{
    \tablefoottext{a}{\cite{vanBreugel_1993}}
    \tablefoottext{b}{\cite{Veron_2010}}
    \tablefoottext{c}{\cite{Baum_1989}}
    \tablefoottext{d}{\cite{Kennicutt_2012}}
    \tablefoottext{e}{\cite{Calzetti_2007}} \\
    The FIR luminosity was computed on the 3-1100 microns range. The $24\: \mu m$ luminosity comes from the WISE archive. The $H\alpha$ luminosity is extinction corrected for 09.6 but not for 3C 285.
  }
\end{table}

   Minkowski Object (MO) is a star-forming peculiar object near the double-lobed FR-\Rmnum{1} radio source NGC 541, in the galaxy cluster Abell 194 \citep{Croft_2006}. In projection, MO is located in a large optical bridge which connects NGC 541 with the interacting galaxies NGC 545/547, suggesting that gas of MO may have origins in previous interactions between these galaxies. MO has recently been observed but not detected in CO(1-0) with the Plateau de Bureau interferometer (Nesvadba et al. in prep, private communication). VLA observations shows the presence of two H\rmnum{1} clouds "wrapped" around the eastern jet with a total mass of $4.9\times 10^8\: M_\odot$ \citep{Croft_2006}. This suggests that H\rmnum{1} may be the result of radiative cooling of warmer gas in the IGM. This highlight a major difference with Centaurus A: in Cen A, the jet probably hits an existing H\rmnum{1} cloud \citep{Mould_2000} whereas the jet of NGC 541 may have caused warm gas to cool forming H\rmnum{1}. General properties of NGC 541 and MO are summarised in table \ref{table:overview_MO}.

\begin{table}[h]
  \centering
  \footnotesize
  \caption{\label{table:overview_MO} General properties of NGC 541 and Minkowski's Object}
  \begin{tabu}{X[l]X[c]X[c]}
    \hline \hline
    Source                                 &           NGC 541           &      Minkowski Object       \\ \hline
    z                                      &       0.0181 \tfm{a}        &       0.0189 \tfm{a}        \\
    $D_A$ (Mpc)                            &            75.8             &            79.1             \\
    $D_L$ (Mpc)                            &            78.6             &            82.1             \\
    Scale (kpc/$''$)                       &            0.37             &            0.38             \\ \hline
    RA (J2000)                             &     $01^h 25^m 44^s.3$      &     $01^h 25^m 47^s.5$      \\
    Dec (J2000)                            &        $-01:22:46.4$        &        $-01:22:20.0$        \\
    $m_B$ (mag)                            &        13.0 \tfm{b}         &        17.5 \tfm{b}         \\ \hline
    $L_{H\alpha}\: (erg.s^{-1})$           & $1.3\times 10^{39}$ \tfm{c} & $6.6\times 10^{40}$ \tfm{a} \\
    $SFR_{H\alpha}\: (M_\odot.yr^{-1})$    &  $7\times 10^{-3}$ \tfm{d}  &        0.35 \tfm{d}         \\
    $L_{24\: \mu m}\: (erg.s^{-1})$        &     $7.4\times 10^{41}$     &     $6.6\times 10^{41}$     \\
    $SFR_{24\: \mu m}\: (M_\odot.yr^{-1})$ &            0.14 \tfm{e}     &        0.13 \tfm{e}         \\
    $L_{FIR}\: (L_\odot)$                  &              -              &  $5.2\times 10^8$ \tfm{f}   \\
    $SFR_{FIR}\: (M_\odot.yr^{-1})$        &              -              &        0.09 \tfm{h}         \\ \hline
  \end{tabu}
  \tablefoot{
    \tablefoottext{a}{\cite{Croft_2006}}
    \tablefoottext{b}{\cite{Simkin_1976}}
    \tablefoottext{c}{\cite{Capetti_2005}}
    \tablefoottext{d}{\cite{Kennicutt_2012}}
    \tablefoottext{e}{\cite{Calzetti_2007}}
    \tablefoottext{f}{\cite{Engelbracht_2008}}
    \tablefoottext{g}{\cite{Kennicutt_1998}} \\
    The $H\alpha$ luminosity is not extinction corrected.
  }
\end{table}

   CO(1-0) and CO(2-1) have been observed along the jet axis of the radio galaxies 3C 285 and NGC 541. Our main driver was to determine the star formation efficiency in the galaxies and inside the jets, and see whether star formation is more efficient in the shocked region along the jet.

   In section \ref{sec:Obs}, we present the data used for this study. The results derived from the observations (section \ref{sec:Res}) are then discussed in section \ref{sec:discussion}. Throughout this paper, we assume the cold dark matter concordance Universe, with $H_0=70 km.s^{-1}.Mpc^{-1}$, $\Omega_m=0.30$ and $\Omega_A=0.70$.

\section{Observations}
\label{sec:Obs}

   \subsection{IRAM-30m, Pico Veleta}

\begin{figure*}[h!]
  \centering
  \includegraphics[width=0.9\linewidth]{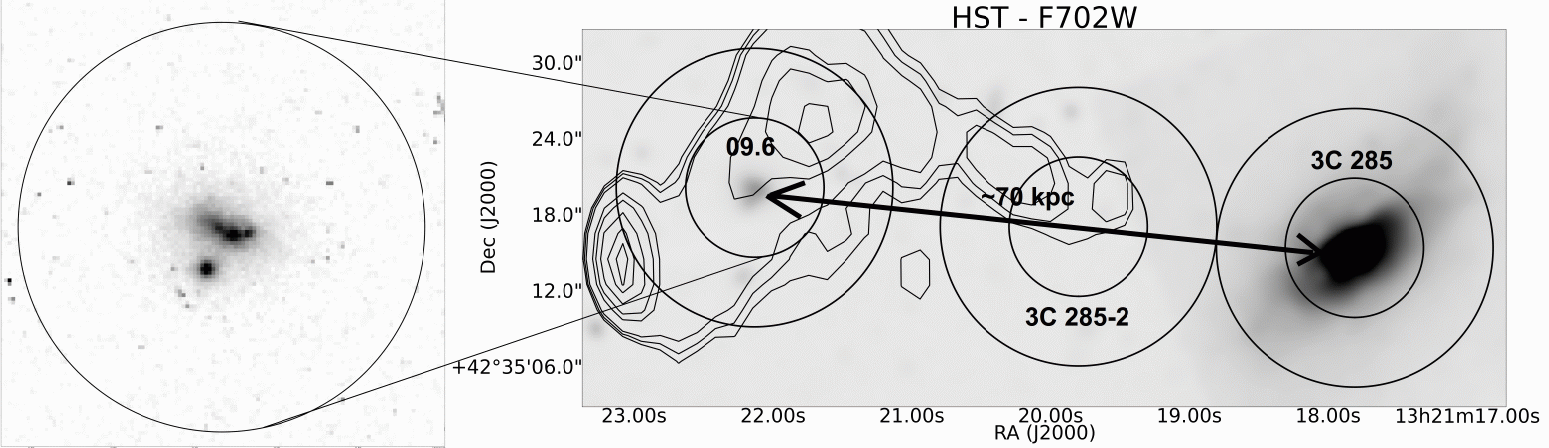}
  \caption{\label{3C285} Contour map of the eastern lobe of 3C 285 observed at 21 cm \citep{vanBreugel_1993} with $5''$ resolution, as extracted from the VLA archive (NED, \citealt{Leahy_1984}), overlaid on a slightly smoothed $H\alpha$ image from HST in the F702W filter (data from the HST archive, PI: Crane). The observed positions are shown by the CO(1-0) $24''$ and CO(2-1) $12''$ IRAM-30m beams (circles). Details of the 09.6 spot are shown in the circle on the left, and show that the spot is resolved in two or maybe three sub-structures.}
\end{figure*}

\begin{figure*}[h!]
  \centering
  \includegraphics[width=0.9\linewidth]{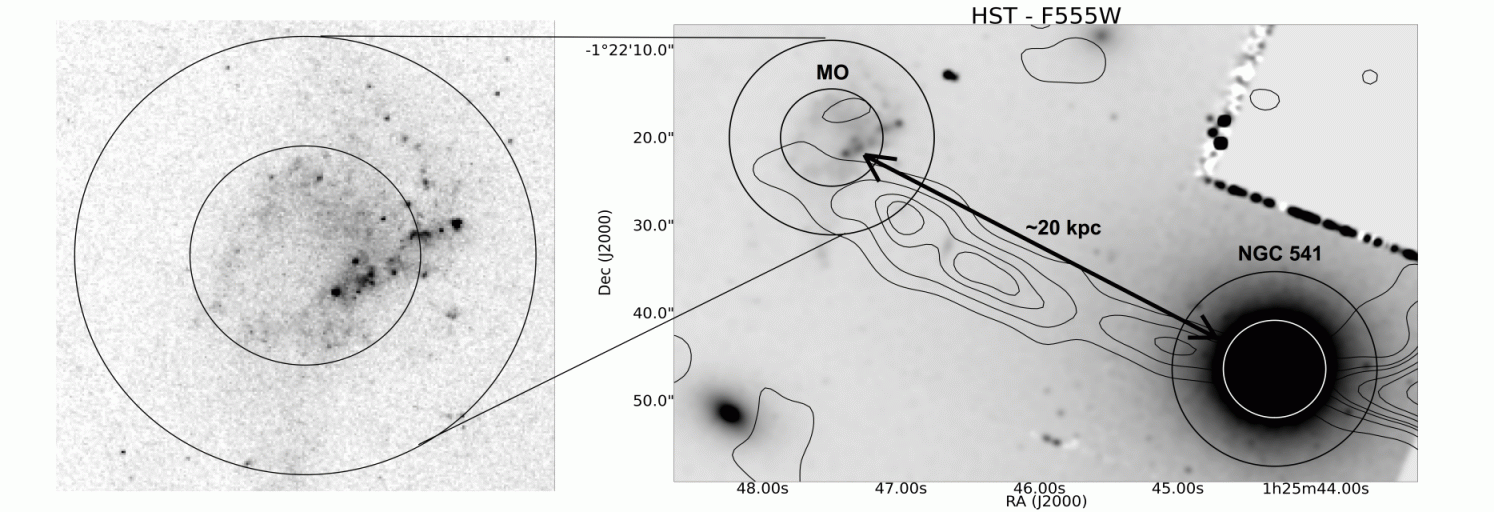}
  \caption{\label{NGC541} Contour map of the eastern lobe of NGC 541 observed at 21 cm \citep{vanBreugel_1985} with $3''$ resolution, as extracted from the VLA archive, overlaid on a slightly smoothed stellar continuum image from HST in the F555W filter (data from the HST archive, PI: Baum). The observed positions are shown by the CO(1-0) $22''$ and CO(2-1) $11''$ IRAM-30m beams (circles). Details of Minkowski Object are shown on the left, and show that the spot is resolved in sub-structures.}
\end{figure*}

   Millimetre observations of the CO(1-0) and CO(2-1) emission were made with the IRAM 30m-telescope on March and June 2014. At redshift z=0.0794 (resp. z=0.0181), those lines are observable at frequencies of 106.780 GHz (resp. 113.220 GHz) and 213.580 GHz (resp. 226.439 GHz), which leads to beams of $24''$ and $12''$ (resp. $22''$ and $11''$). The EMIR receiver were used simultaneously with the 4MHz, FTS and WILMA backends (bandwidths of 4 GHz, 4 GHz and 3.7 GHz; resolution of 4 MHz, 195 kHz and 2 MHz respectively). Due to instrumental problems, the FTS backend could not be used. Moreover, the 4MHz backend was only prepared for the CO(1-0) line. Therefore, only the WILMA backend is used in this paper.

   During observations, the pointing was monitored by observing Mars and standard continuum sources tuned to the frequency corresponding to the redshifted CO(1-0) emission line. Observations were obtained using wobbler switching with a rate of $\sim 0.5\: Hz$. Six-minute scans were taken, and a calibration was done every 3 scans. Pointing was checked every few hours by observing standard continuum sources and was generally determined to be accurate to within a few arcseconds.

   Three regions were observed on March 2014: the central galaxy 3C 285, the 09.6 spot and an intermediate position (3C 285-2) along the jet (cf. figure \ref{3C285}). Along observing nights, the system temperature remained good. It varied between 100 and 165 K for the CO(1-0) line and it was in the range 240-510 K for CO(2-1). On June 2014, observations pointed on NGC 541 and Minkowski Object (MO) (cf. figure \ref{NGC541}), with system temperatures of 180-200 K for CO(1-0) emission and 210-250 K for the CO(2-1) line. In order to reach a better rms and favour the detection, our data of NGC 541 were combined with those of \cite{Ocana_2010}. Observations conditions are summarised in table \ref{table:obs}. \\

\begin{table}[h]
  \centering
  \footnotesize
  \begin{tabular}{lcccc}
    \hline \hline
      Source  &  line   &  frequency  &    $\delta \nu$    &   rms   \\ \hline
      3C 285  & CO(1-0) & 106.780 GHz & $44.9\: km.s^{-1}$ & 0.67 mK \\
              & CO(2-1) & 213.580 GHz &                    & 1.70 mK \\
    09.6 spot & CO(1-0) & 106.780 GHz & $44.9\: km.s^{-1}$ & 0.49 mK \\
              & CO(2-1) & 213.580 GHz &                    & 1.06 mK \\
    3C 285-2  & CO(1-0) & 106.780 GHz & $44.9\: km.s^{-1}$ & 0.78 mK \\
              & CO(2-1) & 213.580 GHz &                    & 1.58 mK \\ \hline
     NGC 541  & CO(1-0) & 113.211 GHz & $42.4\: km.s^{-1}$ & 0.55 mK \\
              & CO(2-1) & 226.417 GHz & $26.5\: km.s^{-1}$ & 1.12 mK \\
        MO    & CO(1-0) & 113.143 GHz & $44.9\: km.s^{-1}$ & 1.01 mK \\
              & CO(2-1) & 226.286 GHz &                    & 1.35 mK \\ \hline
  \end{tabular}
  \caption{\label{table:obs} Journal of observations at IRAM-30m for the project 211-13. The rms were determined with both polarisations. They are given in main beam temperature.}
\end{table}

\begin{figure}[h]
  \centering
  \includegraphics[width=\linewidth]{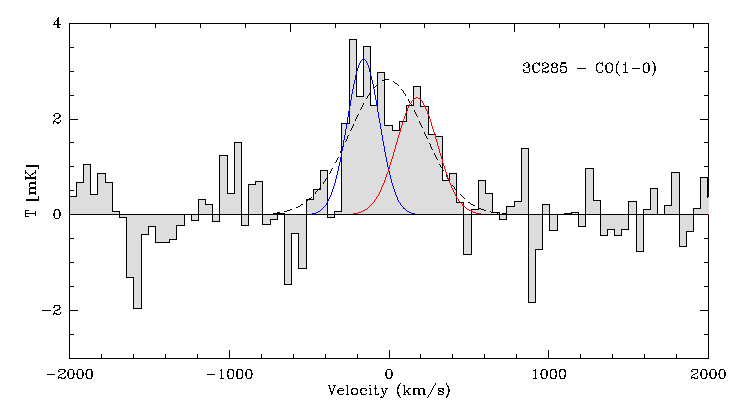}
  \includegraphics[width=\linewidth]{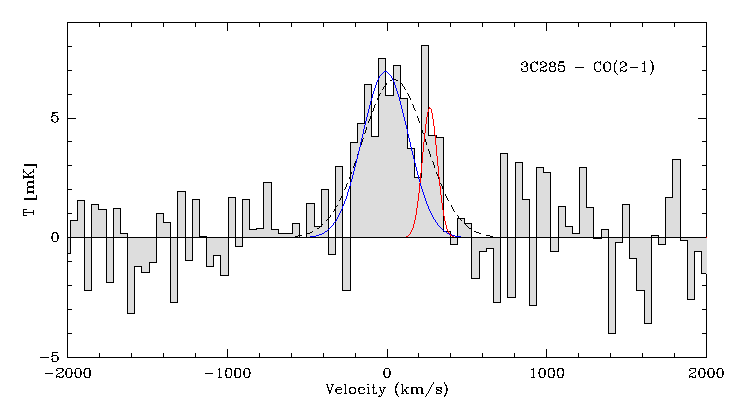}
  \caption{\label{spectra_3C285} \emph{Top:} CO(1-0) spectrum for 3C 285. The dash line fits the total emission; the red and blue lines fit the double-horn profile. \emph{Bottom:} CO(2-1) spectrum for 3C 285. The spectra are tuned to the systemic redshift ($z=0.0794$).}
\end{figure}

\begin{figure}[h]
  \centering
  \includegraphics[width=\linewidth]{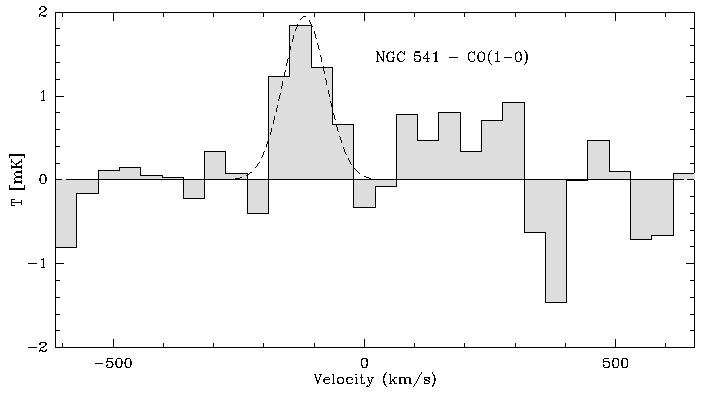}
  \includegraphics[width=\linewidth]{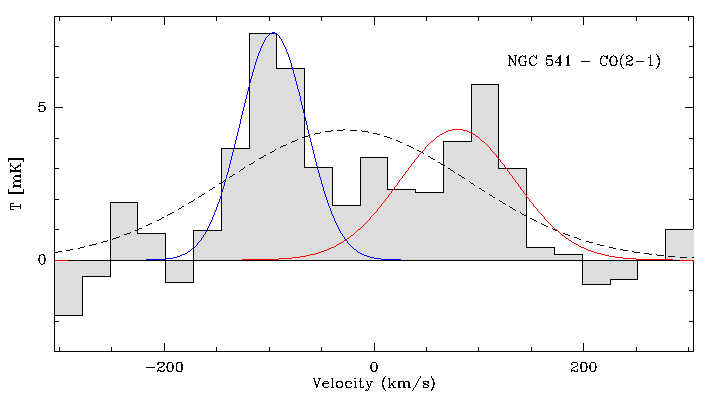}
  \caption{\label{spectra_NGC541} \emph{Top:} CO(1-0) spectrum of NGC 541. The dash line fits the total emission; the red and blue lines fit the double-horn profile. \emph{Bottom:} CO(2-1) spectrum for NGC 541. The spectra are tuned to the systemic redshift ($z=0.0181$).}
\end{figure}

   Data reduction was done using the IRAM package CLASS. After dropping bad spectra, a linear baseline was subtracted from the average spectrum; for detections, the baseline subtraction was done at velocities outside the range of the emission line ($-500$ to $500\: km.s^{-1}$). Then, each spectrum was smoothed to a spectral resolution of $\sim 40-45\: km/s$, except the CO(2-1) spectrum of NGC 541 which has a resolution of $\sim 25\: km/s$. The resultant spectra are plotted in figures \ref{spectra_3C285} and \ref{spectra_NGC541}.\\

   \subsection{Herschel}

\begin{figure*}[ht]
  \centering
  \includegraphics[width=0.8\linewidth]{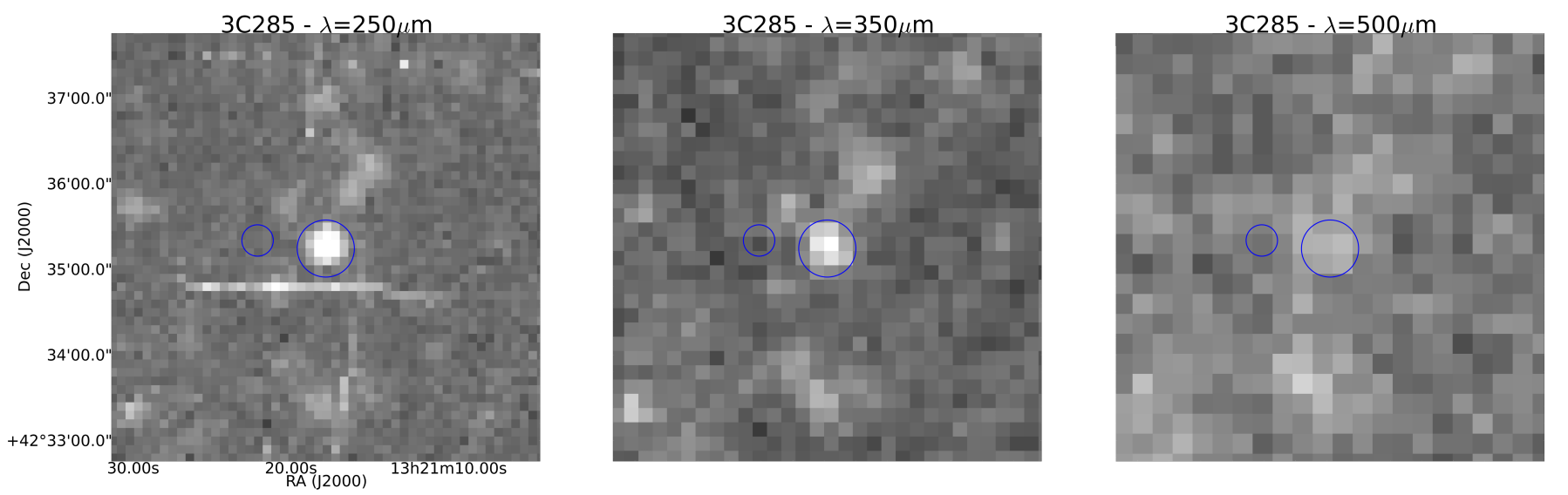}
  \caption{\label{Herschel} Image of the Herschel-SPIRE data of the 3C285 region at 250, 350 and $500\: \mu m$. The central galaxy and the 09.6 spot are shown with blue circles.}
\end{figure*}

   The area around 3C 285 has been mapped with Herschel \citep{Herschel}. The observation were made with the PACS instrument \citep{PACS} at wavelengths $160\: \mu m$ and with the SPIRE instrument \citep{SPIRE} at wavelengths 250, 350 and $500\: \mu m$ (see figure \ref{Herschel}). We used these data, available in the online archive (ObsID: 1342256880). The fluxes of the central galaxy were then extracted (see table \ref{table:Herschel}) before computing the Spectral Energy Distribution (SED). To decrease the number of degrees of liberty, the Spitzer 70 microns flux \citep{Dicken_2010} and the IRAS 60 and 100 microns fluxes were also used.

\begin{table}[h]
  \centering
  \footnotesize
  \begin{tabular}{lcccc}
    \hline \hline
      Source  &       $250\: \mu m$        &       $350\: \mu m$        &       $500\: \mu m$        \\ \hline
      3C 285  &    $1.00\pm 0.013\: Jy$    &    $0.24\pm 0.011\: Jy$    &    $0.14\pm 0.011\: Jy$    \\
    09.6 spot & $<3.79\times 10^{-2}\: Jy$ & $<3.41\times 10^{-2}\: Jy$ & $<3.43\times 10^{-2}\: Jy$ \\ \hline
  \end{tabular}
  \caption{\label{table:Herschel} Fluxes extracted from the Herschel-SPIRE data. For 09.6, an upper limit is given.}
\end{table}

\section{Results}
\label{sec:Res}

   \subsection{CO luminosities}

   The central galaxy 3C 285 was detected in both CO(1-0) and CO(2-1). Each line was fitted by a gaussian in order to get its characteristics. Line fluxes were measured by numerically integrating over the channels in the line profile, and the line widths were measured as full width at 50\% of the peak flux. Then, $L'_{CO}$ was calculated with the formula from \cite{Solomon_1997}. Results are summarised in table \ref{table:spec} and give $L'_{CO}=(2.2\pm 0.2)\times 10^9\: K.km.s^{-1}.pc^2$ for 3C 285.

   For the other positions, there is no detection. Therefore, an upper limit of the line fluxes has been calculated at $3\sigma$ with the line width of the $H\alpha$ emission equal to $64\: km.s^{-1}$ \citep{vanBreugel_1993}. The computed upper limits are $L'_{CO}\leq 1.4\times 10^8\: K.km.s^{-1}.pc^2$ and $L'_{CO}\leq 2.1\times 10^8\: K.km.s^{-1}.pc^2$ for the 09.6 spot and 3C 285-2 respectively.

   The molecular gas mass was estimated from the line luminosity $L'_{CO}$. A standard Milky Way conversion factor of $4.6\: M_\odot.(K.km.s^{-1}. pc^2)^{-1}$ \citep{Solomon_1997} was used to find a molecular gas mass of a few $10^9\: M_\odot$ for the central galaxy whereas the other positions contain less than $10^9\: M_\odot$.

\begin{table*}[h]
  \centering
  \footnotesize
  \begin{tabular}{lccccccc}
    \hline \hline
      Source  &  line   & $T_{mb}$ &   $\Delta v$  &    $v_{peak}$    &     $I_{CO}$    &          $L'_{CO}$          &      $M_{H_2}$     \\
              &         &   (mK)   & ($km.s^{-1}$) &   ($km.s^{-1}$)  & ($K.km.s^{-1}$) & ($10^8\: K.km.s^{-1}.pc^2$) & ($10^8\: M_\odot$) \\ \hline
      3C 285  & CO(1-0) &   2.82   &  $553\pm 51$  &  $-11.6\pm 26.4$ & $1.66\pm 0.15$  &           $22\pm 2$         &     $103\pm 9$     \\
              & CO(2-1) &   6.60   &  $472\pm 55$  &  $44.2\pm 25.5$  & $3.31\pm 0.36$  &              -              &          -         \\ \hline
    09.6 spot & CO(1-0) & $<1.47$  &       -       &        -         &    $<0.100$     &            $<1.4$           &       $<6.2$       \\
              & CO(2-1) & $<3.18$  &       -       &        -         &    $<0.216$     &              -              &          -         \\ \hline
    3C 285-2  & CO(1-0) & $<2.34$  &       -       &        -         &    $<0.159$     &            $<2.1$           &       $<9.9$       \\
              & CO(2-1) & $<4.74$  &       -       &        -         &    $<0.322$     &              -              &          -         \\ \hline
              &         &          &               &                  &                 &                             &                    \\ \hline
     NGC 541  & CO(1-0) &   1.95   &  $101\pm 25$  & $-118.8\pm 12.8$ & $0.21\pm 0.05$  &        $0.14\pm 0.03$       &    $0.6\pm 0.1$    \\
              & CO(2-1) &   4.27   &  $275\pm 27$  &  $-26.5\pm 13.1$ & $1.25\pm 0.12$  &        $0.37\pm 0.04$       &    $1.7\pm 0.2$    \\ \hline
        MO    & CO(1-0) & $<3.03$  &       -       &        -         &    $<0.032$     &            $<0.02$          &       $<0.10$      \\
              & CO(2-1) & $<4.05$  &       -       &        -         &    $<0.043$     &              -              &          -         \\ \hline
  \end{tabular}
  \caption{\label{table:spec} Results of the observations at IRAM-30m. For non-detections, an upper limits is computed at $3\sigma$, with a line width of $\Delta v=64\: km.s^{-1}$ for 09.6 and 3C 295-2 \citep{vanBreugel_1993}, and a line width $\Delta v=10\: km.s^{-1}$ for MO.}
\end{table*}

   \cite{Evans_2005} observed 3C 285 with the NRAO 12m telescope. The CO(1-0) emission line was not detected in the central galaxy. Their non-detection is consistent with our results. It may be explained by the size of the NRAO 12m antenna. which has a main beam about 6 times larger than the IRAM-30m main beam (more dilution effects). The line may also have been subtracted with the baseline.
\bigskip

   NGC 541 was detected in CO(2-1) and partially detected in CO(1-0). The CO(2-1) luminosity of NGC 541 is: $L'_{CO(2-1)}=(3.7\pm 0.4)\times 10^7\: K.km.s^{-1}.pc^2$. For the CO(1-0) emission, only the blueshifted part of the spectrum is used to get a lower limit of the luminosity: $L'_{CO(1-0)}\geq (1.4\pm 0.3)\times 10^7\: K.km.s^{-1}.pc^2$. To calculate the mass of $H_2$, we use the CO(2-1) luminosity assuming a ratio $C0(2-1)/CO(1-0)$ of 2.3.

   For MO, there is no detection and an upper limit at $3\sigma$ has been calculated with a line width of $10\: km.s^{-1}$ estimated from the size and the stellar mass, taken as the dynamical mass. The computed upper limit is $L'_{CO}\leq 2.2\times 10^6\: K.km.s^{-1}.pc^2$. Results are summarised in table \ref{table:spec}.

   The molecular gas mass was estimated with a standard Milky Way conversion factor of $4.6\: M_\odot.(K.km.s^{-1}. pc^2)^{-1}$ \citep{Solomon_1997}. A molecular gas mass of a few $10^8\: M_\odot$ was found for NGC 541 whereas MO contains less than $10^7\: M_\odot$.

   \subsection{Line intensities and line ratios}

   Both CO lines are detected for 3C 285. The peak temperature and the integrated luminosity of the CO(2-1) line are about twice that for CO(1-0). For a point source, if the lines are thermally excited, the brightness temperatures would be similar in the Rayleigh-Jeans regime. Therefore, due to the beam dilution, the line main beam temperature ratio would be about 4. As the CO emission is not resolved (see section \ref{sec:model}), the factor of 2 indicates that the $J=2$ level is subthermally excited, as frequently observed in galaxies \citep{Wiklind_1995}. But we note that the $CO(2-1)/CO(1-0)$ ratio is particularly low and could indicate a rather low density medium.

   For NGC 541, a line ratio might be computed for the blueshifted part of the spectrum. The peak temperature in CO(2-1) is about 4 times greater and the integrated luminosity of the CO(2-1) line about 3 times greater than that for CO(1-0). Contrary to 3C 285, the gas does look to be thermalised in NGC 541.

   \subsection{Line width and morphology}

   For 3C 285 and NGC 541, there is a broad line profile covering negative and positive velocities. This could result from the rotation of the galaxies around their main axes, as suggested by the apparent double-horn profile. Using CLASS, both lines profiles were fitted with gaussians (figures \ref{spectra_3C285} and \ref{spectra_NGC541}).

   The blueshifted line peak temperature is slightly stronger for both frequencies. The double-horn profile of CO(1-0) emission is well defined for 3C 285, whereas it is less obvious in CO(2-1), probably due to the different beam sizes, or more likely because we work in a very small S/N regime. As discussed in section \ref{sec:model}, the CO emission is likely extended on scales $\sim 3.5\: kpc$ ($\sim 2.3''$) in radius. So the smaller CO(2-1) beam could have missed part of the rotating material.

   In addition, the spectra show no kinematic effect of a molecular outflow, at the level of the 30m sensitivity.

   \subsection{SED and IR luminosity}
   \label{sec:SED}

\begin{figure}[h]
  \centering
  \includegraphics[page=1,width=\linewidth,trim=25 10 50 10,clip=true]{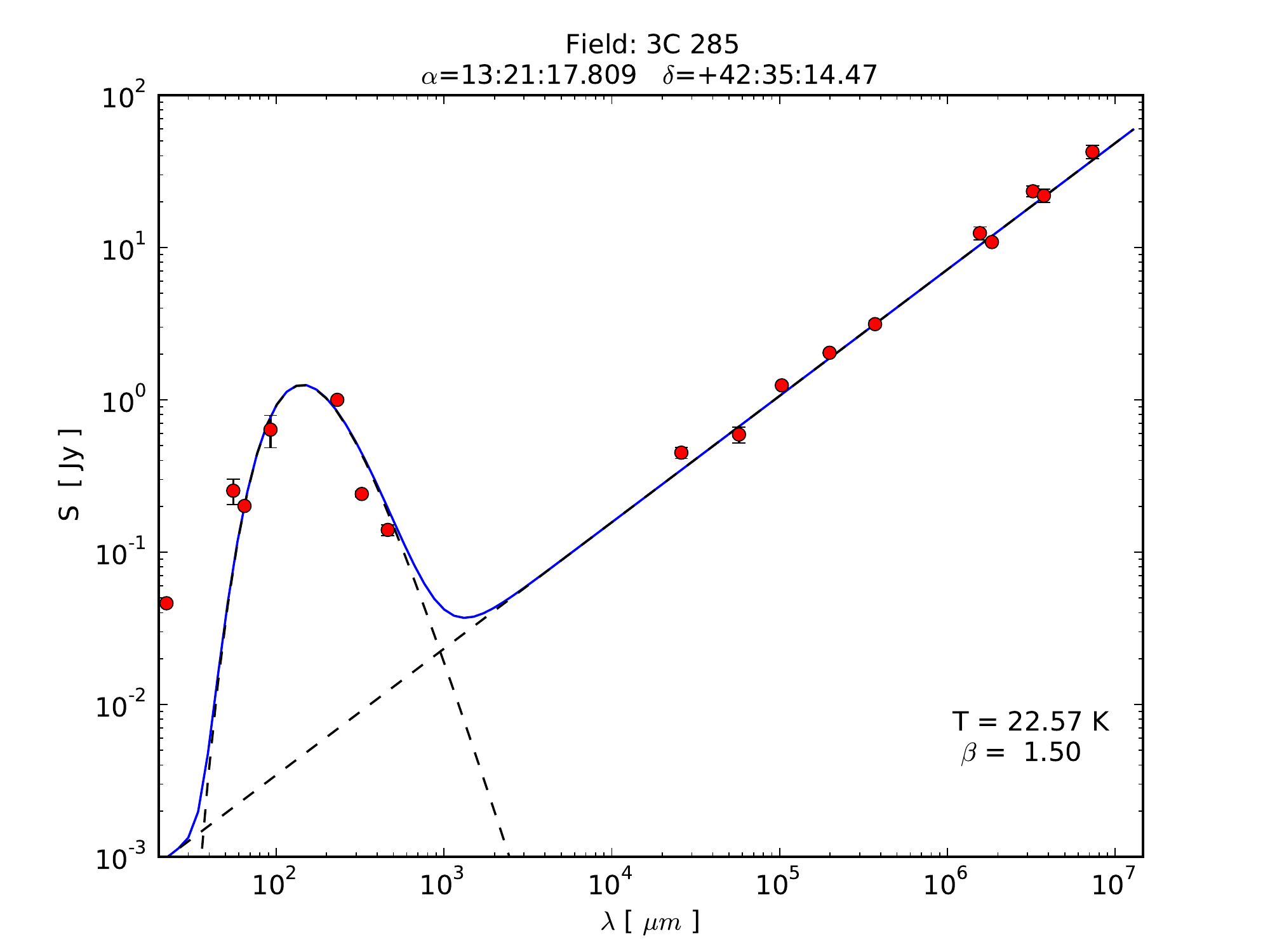}
  \caption{\label{SED} Spectral Energy Distribution of the central galaxy 3C 285. The dash lines represent the synchrotron (power law) and thermal emission (modified black body). The Spitzer 24 microns flux is plotted (not taken into account for fitting), indicating there might be a warm dust component \citep{Dicken_2009}.}
\end{figure}

   We used the Herschel data to fit the SED of 3C 285 and determine its IR luminosity. The IR and radio emissions of the galaxy contain two parts: thermal emission from dust and synchrotron emission due to the AGN. The synchrotron contribution was first fitted by a power law of index about $-0.8$ in frequency with data from literature \citep{Laing_1980, Hales_1988, Gregory_1991, Cohen_2007}. Then, the thermal part of the SED is computed with a fixed $\beta =1.5$ (cf. figure \ref{SED}). It is characterised by a dust temperature $T_{dust}\sim 23\: K$. This temperature gives a dust mass $M_{dust}\sim 2.05\times 10^8\: M_\odot$ \citep{Evans_2005}, which leads to a rather small gas-to-dust ratio $M_{H_2}/M_{dust}\sim 50$. \cite{Wiklind_1995} found that most of the elliptical galaxies have a gas-to-dust ratio of $\sim 700$. However, they also found some ellipticals with a ratio of $\sim 50$. They explain this difference by a lower dust temperature ($<30\: K$), and by the fact that part of the FIR luminosity would come from grains associated with diffuse atomic gas \citep{Wiklind_1995}.

   The TIR luminosity is estimated by integration of the thermal part over the frequencies between 3 and $1100\: \mu m$ (as defined by \citealt{Kennicutt_2012}): $L_{TIR}\approx 1.32\times 10^{11}\: L_\odot$ for 3C 285. This is consistent with the FIR luminosity estimated by \cite{Sanders_1996}: $L_{FIR}=9.51\times 10^{10}\: L_\odot$ and \cite{Satyapal_2005}: $L_{FIR}=7.69\times 10^{10}\: L_\odot$.

   Reported on the $L_{FIR}$ vs $L_{CO}$ diagram \citep{Solomon_1997,Daddi_2010}, 3C 285 seems to be a normal and weakly-interacting galaxy.

   \subsection{Star formation rate}

   The $H\alpha$ emission emerging from photoionised gas by young and massive stars is often used as a tracer of star formation. The $H\alpha$ luminosity can thus be interpreted as a measure of the star formation rate with $SFR=L_{H\alpha}/1.86\times 10^{41}\: erg.s^{-1}$ (see \citealt{Kennicutt_2012} for a review).

   A SFR may also be deduced from the TIR luminosity from dust emission. The emission from young stellar population is partly absorbed by dust which heats and emits in TIR via thermal emission. The relationship between the TIR luminosity and the SFR is given by $SFR=L_{TIR}/6.7\times 10^9\: L_\odot$ \citep{Kennicutt_2012}.

   For IR emission, a SFR can also be derived from the FIR or the $24\: \mu m$ emission: $SFR=L_{FIR}/5.8\times 10^9\: L_\odot$ \citep{Kennicutt_1998} and $SFR=1.27\times 10^{-37}[L_{24\: \mu m} (erg.s^{-1})]^{0.8850}$ \citep{Calzetti_2007}. The $24\: \mu m$ luminosity is calculated with the WISE \citep{WISE} data from the online archive. Table \ref{table:SFR} summarises the different single-wavelength SFR.

\begin{table}[h]
  \centering
  \footnotesize
  \begin{tabular}{lcccc}
    \hline \hline
      Source             & 3C 285 & 09.6 spot &     NGC 541       &   MO    \\ \hline
      $SFR_{H\alpha}$    &  1.34  &   0.15    & $7\times 10^{-3}$ &  0.35   \\
      $SFR_{TIR}$        &  19.7  &  $<0.79$  &         -         &    -    \\
      $SFR_{FIR}$        &   -    &     -     &         -         &  0.09   \\
      $SFR_{24\: \mu m}$ &  12.0  &   0.61    &       0.14        &  0.13   \\ \hline
  \end{tabular}
  \caption{\label{table:SFR} SFR in $M_\odot.yr^{-1}$ for the different objects observed with the IRAM-30m. The $H\alpha$, TIR, FIR and $24\: \mu m$ SFR were calculated with \cite{Kennicutt_2012}, \cite{Kennicutt_1998} and \cite{Calzetti_2007}. The total SFR is a combination of those SFR.}
\end{table}

   The values of SFR differ from each other with factors up to 20. This may be explained by dust absorption. Indeed, the optical image of 3C 285 shows a dust lane able to obscure the $H\alpha$. Therefore, the SFR derived from the $H\alpha$ emission is underestimated. On the opposite, the 09.6 spot, NGC 541 and MO show little dust absorption.

   The other SFR estimations are also contaminated. Contrary to the $H\alpha$ emission that traces the young and massive stars, the IR emission traces the dust thermal emission after absorption of the starlight. Moreover, the IR emission also contains a contribution from old stars. Therefore, the SFR derived from the IR emission is overestimated.

   To solve this problem, we used a multiwavelength estimation of the SFR. We corrected the $H\alpha$ emission with the $24\: \mu m$ emission using a formula from \cite{Kennicutt_2012}: $L_{H\alpha}^{corr}=L_{H\alpha}^{obs}+0.020\: L_{24\: \mu m}$. The SFR is then calculated from the corrected $H\alpha$ emission using the conversion formula above. Table \ref{table:new_SFR} summarises the dust-attenuation correction of $H\alpha$ and the corresponding SFR.

\begin{table}[h]
  \centering
  \footnotesize
  \begin{tabular}{lcccc}
    \hline \hline
      Source                                        & 3C 285 & 09.6 spot & NGC 541 &  MO  \\ \hline
      $L_{H\alpha}^{obs}\: (10^{40}\: erg.s^{-1})$  &   25   &    2.8    &  0.13   & 6.6  \\
      $L_{24\: \mu m}\: (10^{42}\: erg.s^{-1})$     &  110   &    3.8    &  0.74   & 0.66 \\
      $L_{H\alpha}^{corr}\: (10^{40}\: erg.s^{-1})$ &  245   &   10.4    &  1.61   & 7.92 \\ \hline
      $SFR\: (M_\odot.yr^{-1})$                     & 14.53  &   0.62    &  0.095  & 0.47 \\
      $t_{dep}\: (Gyr)$                             &  0.71  &  $<1.0$   &  1.79   & 0.02 \\ \hline
   \end{tabular}
   \caption{\label{table:new_SFR} Dust-attenuation corrections of the $H\alpha$ emission using the formula from \cite{Kennicutt_2012} for the different objects observed with the IRAM-30m. The SFR are calculated with the conversion formula \citep{Kennicutt_2012}.}
\end{table}

   The depletion time is the time to consume all the gas with the present star formation rate: $t_{depl}\sim M_{gas}/SFR$. For 3C 285, the CO(1-0) emission line gives a gas mass $M_{H_2}=(10.3\pm 0.9)\times 10^9\: M_\odot$ with the standard conversion factor. Therefore the depletion time is $(7.09\pm 0.62)\times 10^8\: yr$. As the gas mass of the 09.6 spot is $<6.2\times 10^8\: M_\odot$, one gets a depletion time $<1.0\times 10^9\: yr$. \\
For NGC 541, the gas mass is given by CO(2-1) emission $M_{H_2}=(1.7\pm 0.2)\times 10^8\: M_\odot$ therefore, the depletion time is $(1.79\pm 0.21)\times 10^9\: yr$. The mass of MO is $<1.0\times 10^7\: M_\odot$ so the depletion time is $<2.1\times 10^7\: yr$. The table \ref{table:masses} summarises these results.

\section{Discussion}
\label{sec:discussion}

   \subsection{A Kennicutt-Schmidt law?}

   The gas mass determined from the CO data may now be used to estimate the $M_{H_2}/M_*$ ratio in each object. \cite{Croft_2006} gives a stellar mass of $1.9\times 10^7\: M_\odot$ for MO, and \cite{Tadhunter_2011} gives $M_*\sim 4.2\times 10^{11}\: M_\odot$ for 3C 285. For NGC 541 and 09.6, the stellar masses were calculated with the optical magnitudes and the mass-to-light ratios given by \cite{Bell_2001}. Those masses are summarised in table \ref{table:masses}.

\begin{table}[h]
  \centering
  \footnotesize
  \begin{tabular}{lcccc}
    \hline \hline
                                  & NGC 541 & 3C 285 & 09.6 spot &     MO      \\ \hline
     $M_{gas}\: (10^8\: M_\odot)$ &   1.7   &  103   &  $<6.2$   &   $<0.1$    \\
       $M_*\: (10^9\: M_\odot)$   &   470   &  420   &    1.9    &    0.019    \\
              $f_{gas}$           & 0.0004  & 0.025  & $<0.326$  &   $<0.53$   \\
      $SFR\: (M_\odot.yr^{-1})$   &  0.095  & 14.53  &   0.62    &    0.47     \\
          $t_{dep}\: (Gyr)$       &  1.79   &  0.71  &  $<1.0$   &   $<0.02$   \\
    $sSFR\: (10^{-2}\: Gyr^{-1})$ &  0.02   &  3.5   &    33     & $\sim 2500$ \\ \hline
  \end{tabular}
  \caption{\label{table:masses} Molecular gas, stellar masses and gas fraction for the different objects observed with the IRAM-30m. The stellar masses are calculated with the mass-to-light ratio \citep{Bell_2001}. The depletion time is computed from the CO-derived gas masses and the SFR.}
\end{table}

   3C 285 have a molecular gas to stellar mass ratio of a few percent whereas NGC 541 is very poor in gas. For MO and 09.6, we only have upper limits as shown in table \ref{table:masses}.

   We conclude that 3C 285 lies on the main sequence of galaxies, whereas NGC 541 is a red sequence galaxy. This could indicate that a minor merger happened with 3C 285 and not with NGC 541. The 09.6 spot lies on or over the main sequence. Finally, MO is a peculiar object as it lies above the main sequence.

   Note that we did not take into account the H\rmnum{1} mass. Elliptical galaxies typically have a H\rmnum{1}-to-$H_2$ ratio of 2.5 \citep{Wiklind_1995}. This would lead to a total gas fraction of $\sim 6\%$ in 3C 285. For 3C 285, no upper limit in H\rmnum{1} has been published yet. \cite{Morganti_2005} looked for H\rmnum{1} absorption line in front of the 3C 285 continuum but found neither emission nor absorption. On the opposite, MO has a lot of atomic gas: $M_{H\rmnum{1}}=4.9\times 10^8\: M_\odot$ \citep{Croft_2006}.
\medskip

   We have calculated the gas and SFR surface densities ($\Sigma_{gas}$, $\Sigma_{SFR}$). For 3C 285, both quantities are smoothed over the CO(1-0) IRAM-30m beam, which gives $\Sigma_{H_2}\approx 58.0\pm 5.1\: M_\odot.pc^{-2}$ and $\Sigma_{SFR}\approx 0.082\: M_\odot.yr^{-1}.kpc^{-2}$. For the 09.6 spot, the surface densities are estimated on the area of the $H\alpha$ emission (in a radius of $\sim 1.65''$), giving $\Sigma_{H_2}<179\: M_\odot.pc^{-2}$ and $\Sigma_{SFR}\approx 0.179\: M_\odot.yr^{-1}.kpc^{-2}$.

\begin{figure}[h]
  \centering
  \includegraphics[width=\linewidth,trim=20 0 40 30,clip=true]{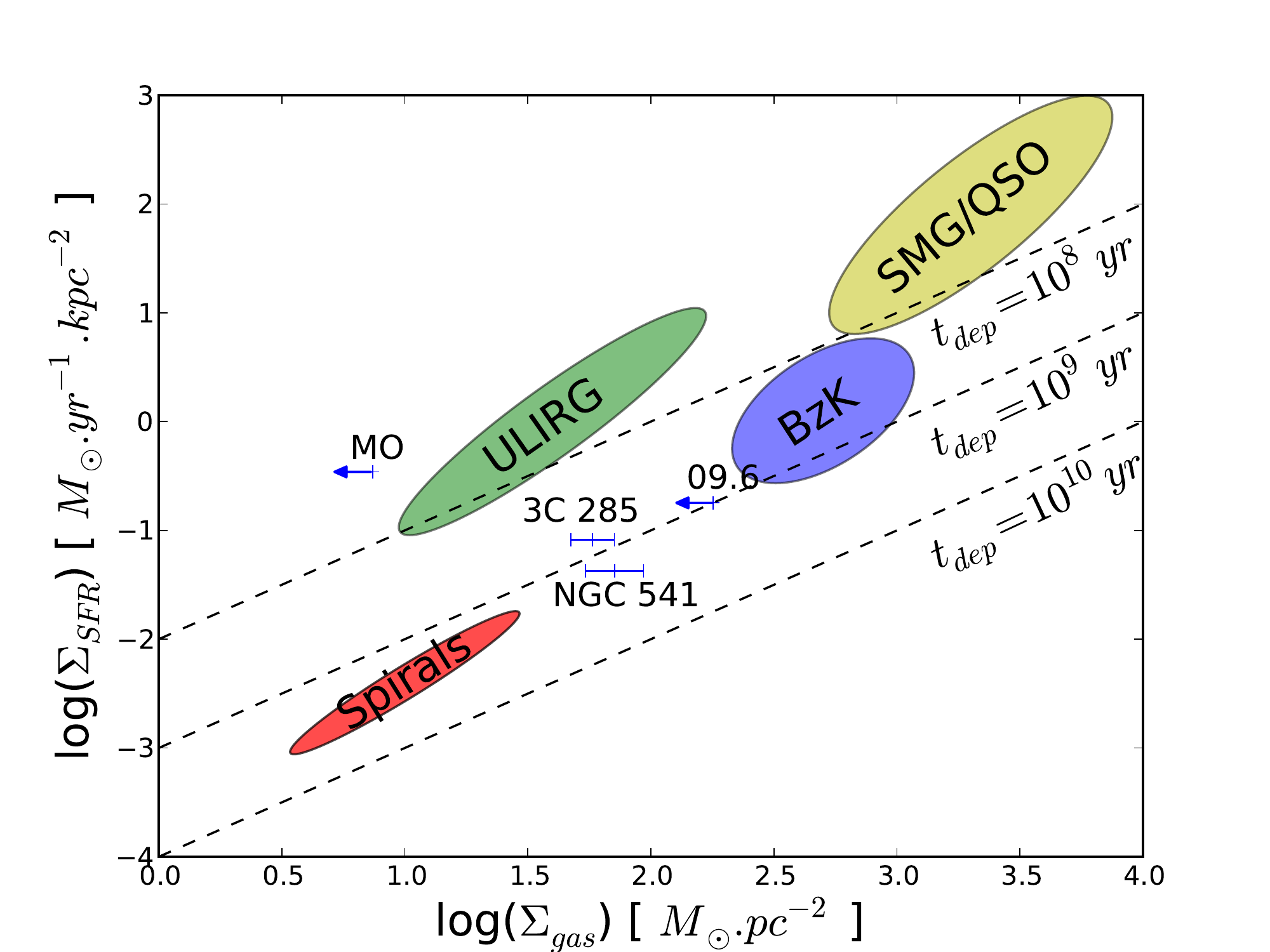}
  \caption{\label{KS-law} $\Sigma_{SFR}$ vs. $\Sigma_{gas}$ for the four sources of this paper. The diagonal dashed lines show lines of constant SF efficiency, indicating the level of $\Sigma_{SFR}$ needed to consume 1\%, 10\%, and 100\% of the gas reservoir in $10^8$ years. Thus, the lines also correspond to constant gas depletion times of, from top to bottom, $10^8$, $10^9$, and $10^{10}\: yr$ The coloured regions come from \cite{Daddi_2010}.}
\end{figure}

   Plotting this in the $\Sigma_{SFR}$ vs $\Sigma_{gas}$ diagram (see figure \ref{KS-law}, \citealt{Bigiel_2008,Daddi_2010}), both positions follow a Schmidt-Kennicutt law $\Sigma_{SFR}\propto \Sigma_{H_2}^N$ \citep{Kennicutt_1998}. To accurately determine the SFE in the 09.6 spot, high-resolution interferometric data are required.

   In NGC 541, the gas and SFR surface densities are smoothed over the CO(2-1) IRAM-30m beam, which gives $\Sigma_{H_2}\approx 71.2\pm 8.4\: M_\odot.pc^{-2}$ and $\Sigma_{SFR}\approx 4.26\times 10^{-2}\: M_\odot.yr^{-1}.kpc^{-2}$. For MO, the surface densities are estimated on the area of the stellar emission (in a radius of $\sim 4.08''$), giving $\Sigma_{H_2}<7.4\: M_\odot.pc^{-2}$ and $\Sigma_{SFR}\approx 0.345\: M_\odot.yr^{-1}. kpc^{-2}$.

   As a conclusion, figure \ref{KS-law} shows that the two star forming regions 09.6 and MO (lying along the AGN radio-jets of 3C285 and NGC541) must have depletion time scales of the order or shorter than typical spiral galaxies. This support the AGN positive feedback scenario that predicts an enhanced star formation activity along the shocked region inside the radio-jets. More sensitive observations to detect and map the CO emission are however necessary to measure accurately the effect of the jet on the gas, its impact on the cooling and thus on the triggered star formation efficiency.

   \subsection{A model to compute velocity spectra}
   \label{sec:model}

   In order to interpret the kinematics of our CO data, we used a simple analytical model which computes the velocity spectrum from the rotation velocity profile of the galaxy \citep{Wiklind_1997}. The rotation velocity profile is determined from the stellar mass distribution, assuming it follows a Plumer distribution.

   The gas distribution was assumed to be axisymmetric of surface density $n(r)$ in a disc with negligible thickness. For each velocity dv, the code calculates the density contained in the isovelocity (equation \ref{eq:dndv}). The velocity spectrum corresponds to the histogram of the velocities. To take into account the gas dispersion, the computed spectrum was then convolved with a gaussian of standard $\sigma=10\: km.s^{-1}$.
\begin{equation} \label{eq:dndv}
  \frac{dN}{dv}(v)=\int \frac{n(r)\, rdr}{v_{rot}(r)\sin i \cbra{1-\frac{v}{v_{rot}(r)\sin i}}^2}
\end{equation}

\begin{figure}[h]
  \centering
  \includegraphics[width=0.49\linewidth]{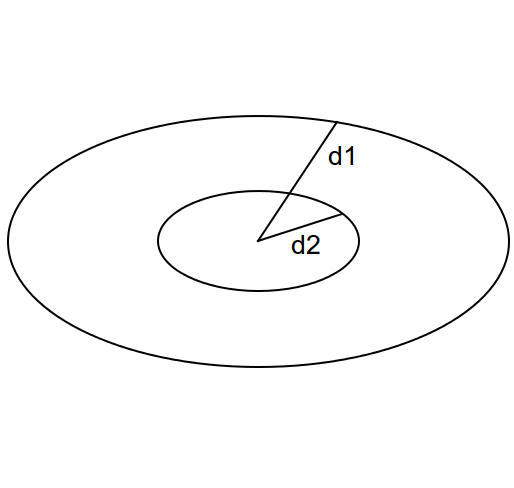}
  \includegraphics[width=0.49\linewidth]{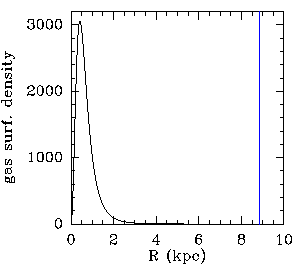}
  \caption{\label{sketch_ring} \emph{Left:} Sketch of a ring with the characteristic distances $d_1$ and $d_2$. \emph{Right:} Surface density profile as derived by the model for a Toomre ring of mass $1.03\times 10^{10}\: M_\odot$ with $d_1=0.9\: kpc$ and $d_2=0.5\: kpc$. The vertical blue line represents the size of the CO(2-1) beam for 3C 285.}
\end{figure}

   We wanted to study the concentration of gas in the galaxy, and determine if the gas is distributed in a disc or a ring. The disc was modelled with a Toomre disc of order 2: $n(r)=n_0\cbra{1+\frac{r^2}{d^2}}^{-5/2}$ \citep{Toomre_1964}. A ring is the difference between two Toomre discs (see sketch in figure \ref{sketch_ring}).

   We ran grids of model, varying the distances $d_1$ and $d_2$. $d_1$ varies from a few hundred parsecs to 10 kpc, and $d_2$ varies from 0 to a few tens parsecs below $d_1$ ($d_2=0$ corresponds to a disc). Both $d_1$ and $d_2$ have influence on the morphology of gas. For small values of $d_1$, the gas is distributed in a narrow dense ring, whereas for larger distances, the ring is broad with broadness of a few kiloparsecs. In addition, for inner radii larger than $\sim 2\: kpc$, the gas ring extends far enough to be slightly resolved by the IRAM 30m-telescope. \\
The inclination angle will also influence the spectra characteristics. As the radial velocity is proportional to the sine of the angle, the peaks go closer to each other as the inclination decreases. The depth does not change significantly with inclination, except for very low angles, when the peaks start overlapping.

   \subsection{A compact molecular ring in 3C 285 and NGC 541}

\begin{figure}[h]
  \centering
  \includegraphics[width=\linewidth]{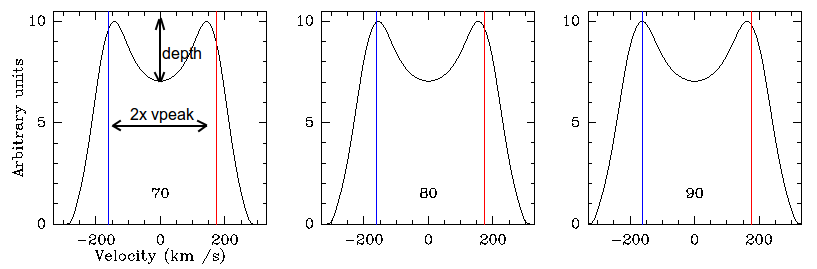}
  \caption{\label{model_3C285} Velocity spectra computed from the model for inclination angle of 70\degree, 80\degree and 90\degree. The vertical blue and red lines indicate the position of the double-horned profile peaks in the data.}
\end{figure}

   \cite{Roche_2000} used V- and R-band data to investigate the radial profiles of radio galaxies. For 3C 285, the best fitting model gives a half-light radius of $\sim 8.3\: kpc$, with a stellar mass of $\sim 4.2\times 10^{11}\: M_\odot$. To compare the models with the data, we used the peak velocity and the relative well depth (see in figure \ref{model_3C285}).

\begin{figure*}[h]
  \centering
  \includegraphics[height=6cm,trim=70 0 100 10,clip=true]{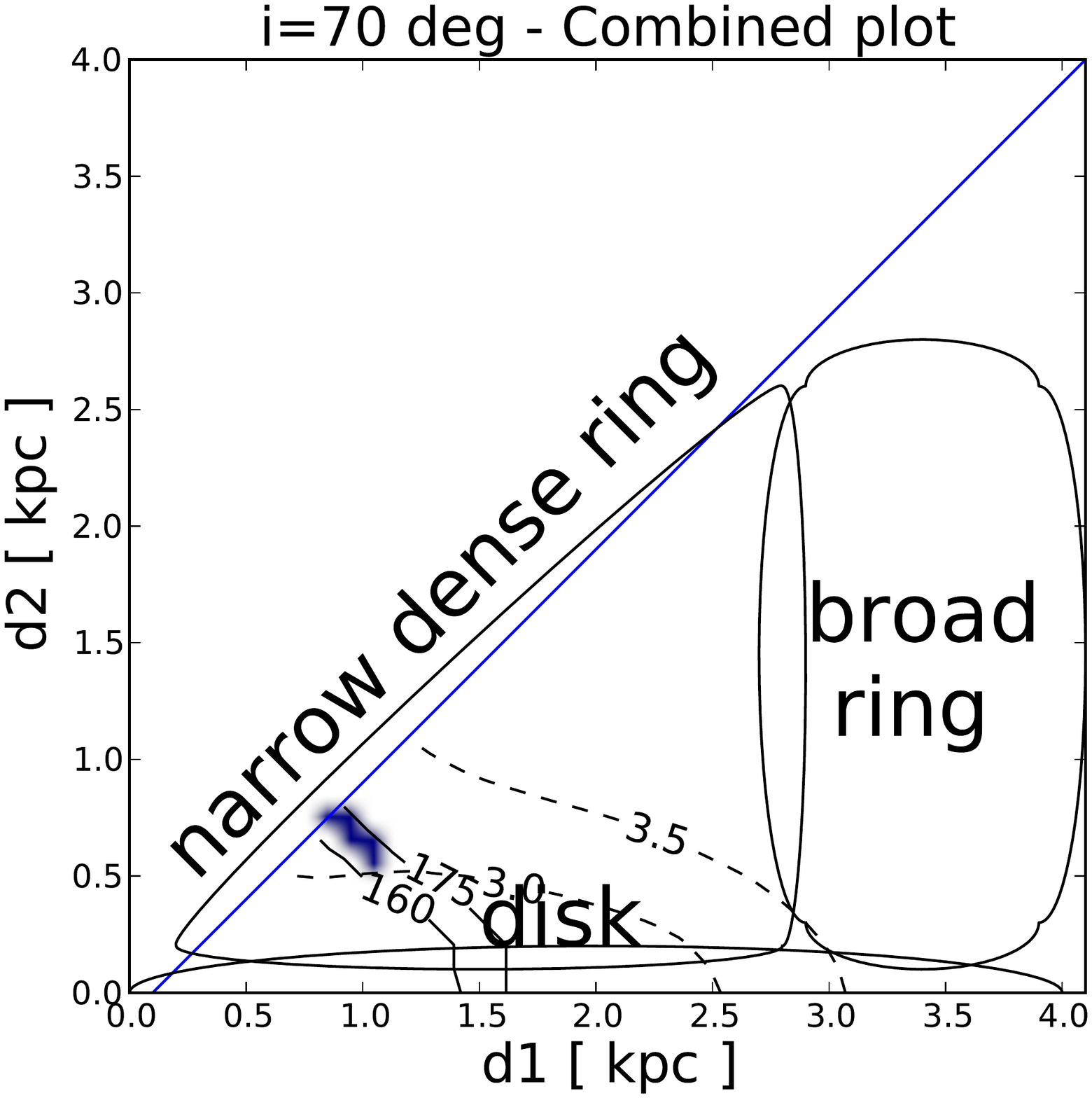}
  \includegraphics[height=6cm,trim=0 0 100 10,clip=true]{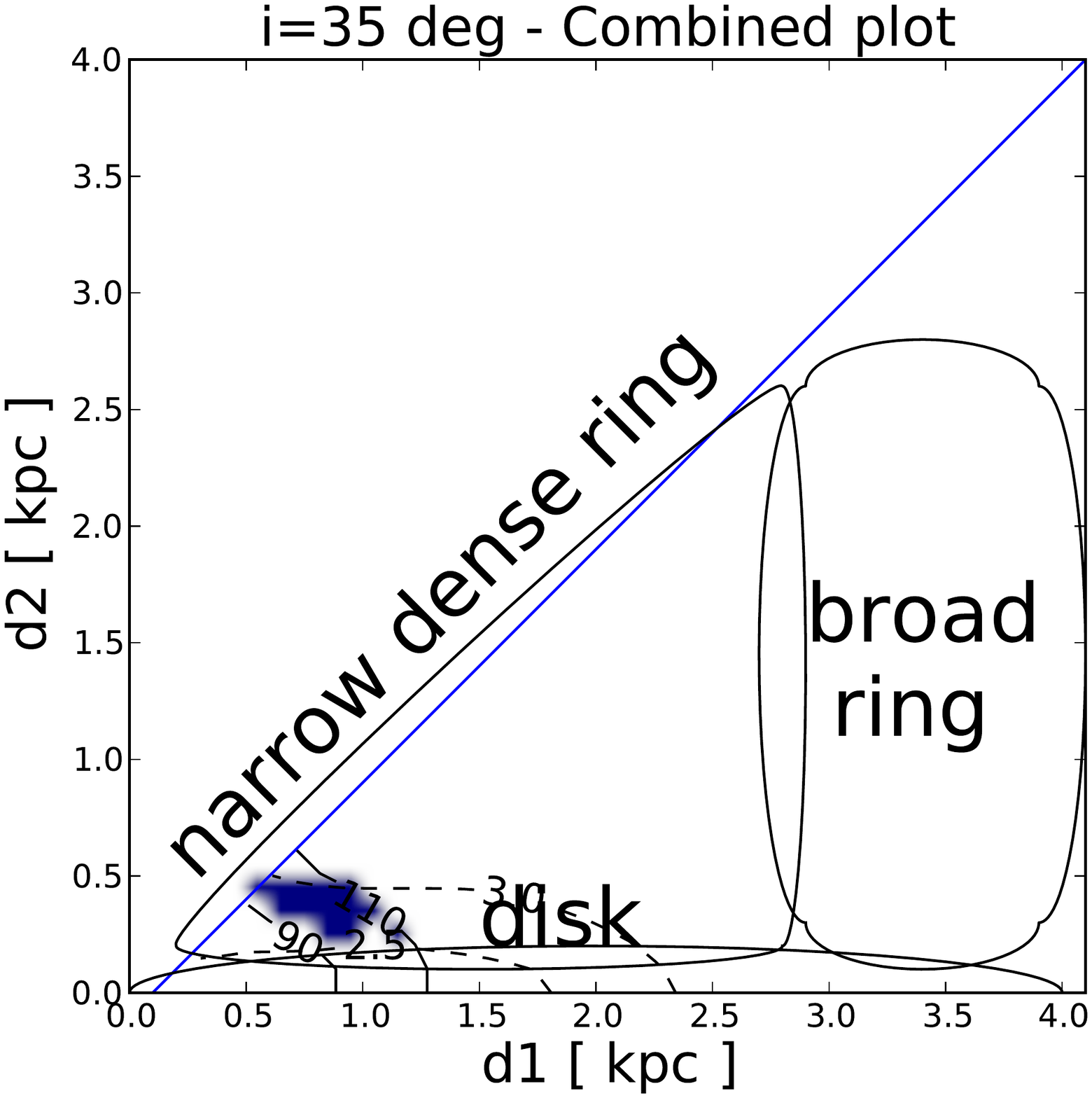}
  \caption{\label{model_results} Parameter space of the model grids for 3C 285 (\emph{left}) and NGC 541 (\emph{right}). The axes indicate the value of the outer $d_1$ and inner $d_2$ distances of the Toomre ring (see sketch in figure \ref{sketch_ring}). Since $d_1>d_2$, the explored parameters lie below the blue line (bottom-right part of the plot). The black solid lines show solutions for a peak velocity of 160 and $175\: km.s^{-1}$ (left), and 90 and $110\: km.s^{-1}$ (right). The black dashed lines show the solutions for a well depth (see figure \ref{model_3C285}) of 30\% and 35\% of the peak (left), and 25\% and 30\% of the peak (right). These values correspond to the observational constraints extracted from the 30m-telescope spectra. The dark blue surface is the intersection that fits both solutions.}
\end{figure*}

   The observational spectrum presents peak velocities of $\sim 160-175\: km.s^{-1}$ and a relative well depth of 1/3. The range of parameters that fit the observations are $d_1=0.7-1.0\: kpc$ and $d_2=0.5-0.7\: kpc$ for an inclination angle larger than 70\degree (see left panel of figure \ref{model_results}), which is consistent with the optical image. Figure \ref{model_3C285} represents the spectra for a ring of $1.03\times 10^{10}\: M_\odot$ with $d_1=0.9\: kpc$ and $d_2=0.5\: kpc$. The density profile (figure \ref{sketch_ring}) indicates that the gas is distributed in a narrow ring that extends at distances up to $\sim 2\: kpc$ with an average radius $\sim 0.7\: kpc$, but this needs to be confirmed by interferometric data.

\smallskip

   For NGC 541, the half-light radius is $\sim 5-8\: kpc$ \citep{Loubser_2012}, with a stellar mass of $\sim 4.7\times 10^{11}\: M_\odot$ \citep{Bell_2001}.

   The observational spectrum presents a peak velocity of $\sim 100\: km.s^{-1}$ and a relative well depth of $1/4-1/3$. The range of parameters that fit the observations are $d_1=0.4-1.1\: kpc$ and $d_2=0.2-0.4\: kpc$ for an inclination angle between 30\degree and 40\degree (see right panel of figure \ref{model_results}), which is consistent with the optical image. The gas is distributed in a narrow ring that extends at distances up to $\sim 2\: kpc$ with an average radius $\sim 0.5\: kpc$, but this also needs to be confirmed by interferometric data.
\medskip

\section{Conclusions}

   We used the IRAM 30m-telescope to observe the centre region of two radio-galaxies: 3C 285 and NGC 541. We also pointed towards two star-forming regions standing along the radio-jet direction of each of these objects: 09.6 at a distance of $\sim 70\: kpc$ from 3C 285 and Minkowski Object (MO) at $\sim 20\: kpc$ from NGC 541.

   NGC 541 was detected in CO(2-1). The CO(1-0) emission line is marginally visible when our observations are combined with the non-detection from \cite{Ocana_2010} on the same object. \\
We derived a total molecular gas mass of $\sim 10^8\: M_\odot$, leading to a gas fraction smaller than 1\%. With a very small star formation rate, this object thus appears like a typical read and dead galaxy. However its depletion timescale ($\sim 2\: Gyr$) is typical of normal star forming galaxies. So the very small star formation rate is mostly due to the small gas fraction in this object. The CO line profile has a typical double-horn shape that indicates a possible rotating disk or ring. In order to reproduce the molecular gas velocity profile, we ran simplified analytical models, constrained by estimates of the stellar mass of the system, its effective radius and its inclination. These models managed to reproduce the CO line profiles with a rather compact ring-like distribution ($\sim 1-2\: kpc$). \\
The origin of this gas is not discussed here, but as already deduced for other objects like Centaurus A, rotating rings of molecular gas are the expected remnants of a recent minor merger activity.

   3C 285 was detected in CO(1-0) and CO(2-1) with a total molecular gas mass of $\sim 10^{10}\: M_\odot$, meaning a gas fraction of $\sim 2.5\%$. Surprisingly for this kind of source, 3C 285 has a fairly high SFR of $\sim 15\: M_\odot.yr^{-1}$. With a depletion time of less than one Gyr, this source looks like typical star forming galaxies, as shown by its position in a KS-diagram. The simple analytical models that reproduce the CO line profiles are also the ones where the gas is lying in a compact molecular ring/disk of 1-2 kpc. So star formation may proceed in this very compact region hidden inside a larger dust-lane seen in the optical image and which crosses the entire galaxy. Note that 3C 285 has a much more massive molecular gas reservoir than NGC 541, standing in a disk of about the same size. So the molecular gas density must be higher which could explain why the SFE ($1/t_{dep}$) is higher in this object.

   MO and 09.6 have not been detected in CO by the 30m-telescope. However, we reached interesting rms for these two sources, leading to upper limits of molecular gas amount of $\sim 10^7-10^8\: M_\odot$. This means that 09.6 and MO have a depletion time of $\le 1\: Gyr$ and $\le 0.02\: Gyr$ respectively. In a KS-diagram, 09.6 lies with or above the normal star forming galaxies and MO with the highly efficient star forming objects. This result shows that the star formation observed in the radio-lobes of 3C 285 and NGC 541 is at least as efficient as inside spiral galaxies and even boosted in the case of the MO.

   If present as suggested by the star formation activity, the origin of the molecular gas in 09.6 and MO is still an open question. However, the differences in the molecular-to-atomic gas fraction, the gas-to-dust ratio or the specific star formation rate in these two objects indicate different scenarios. 09.6 could be the remnant of a small galaxy that has lost most of its gas in a tidal interaction and that is being compressed by the interaction with the 3C 285 radio-lobe. In the case of MO, the presence of atomic gas, the short depletion time scale and the very high sSFR may indicate a recent star forming event that has not produced many stars yet. While 09.6 has a stellar mass of $\sim 10^9\: M_\odot$, typical of a small galaxy, the MO is 100 time smaller with a stellar mass that only reaches $\sim 10^7\: M_\odot$. The small amounts of molecular gas in MO could be explained if the gas is mainly atomic or if the metallicity is too low to keep the standard conversion factor valid. In such a case, the MO could have condensed, after the interaction with the radio-lobes, from the low-metallicity intergalactic medium surrounding NGC 541, as already suggested in the filaments of several brightest cluster galaxies.

   This is consistent with the modelling done by \cite{Fragile_2004} where the authors applied their hydrodynamic simulations of radiative shock-cloud interactions to MO as a test case. They concluded that MO could result from an interaction of a $\sim 10^5\: km.s^{-1}$ jet with an ensemble of moderately dense ($10\: cm^{-3}$) and warm ($10^4\: K$) intergalactic clouds, the large H\rmnum{1} mass in MO being explained by the radio-jet triggered radiative cooling of the warm surrounding gas.

\begin{acknowledgements}
   We thank the anonymous referee for his useful remarks. We also thank Santiago Garcia-Burillo for helpful comments. \\

   Herschel is an ESA space observatory with science instruments provided by European-led Principal Investigator consortia and with important participation from NASA. \\

   This research has made use of the NASA/IPAC Extragalactic Database (NED) which is operated by the Jet Propulsion Laboratory, California Institute of Technology, under contract with the National Aeronautics and Space Administration. \\

   This research has made use of the NASA/IPAC Infrared Science Archive, which is operated by the Jet Propulsion Laboratory, California Institute of Technology, under contract with the National Aeronautics and Space Administration. \\

   This publication makes use of data products from the Wide-field Infrared Survey Explorer, which is a joint project of the University of California, Los Angeles, and the Jet Propulsion Laboratory/California Institute of Technology, funded by the National Aeronautics and Space Administration. \\

   F.C. acknowledges the European Research Council for the Advanced Grant Program Number 267399-Momentum.
\end{acknowledgements}

\bibliography{Biblio}
\bibliographystyle{aa}

\end{document}